\documentclass{article}
\usepackage{amssymb}
\usepackage{graphicx}
\usepackage{amsmath}

\input{tcilatex}

\begin{document}

\title{Weak pseudo-Hermiticity and antilinear commutant\thanks{%
Supported in part by PRIN 2000 ''Sintesi''.}}
\author{L. Solombrino \\
Dipartimento di Fisica dell'Universit\`{a} di Lecce\\
and INFN, Sezione di Lecce, I-73100 Lecce, Italy\thanks{
e-mail: luigi.solombrino@le.infn.it}}
\maketitle

\begin{abstract}
We inquire into some properties of diagonalizable pseudo-Hermitian
operators, showing that their definition can be relaxed and that the
pseudo-Hermiticity property is strictly connected with the existence of an
antilinear symmetry. This result is then illustrated by considering the
particular case of the complex Morse potential.


\newpage
\end{abstract}

\section{Introduction}

In the last years the study of some non-Hermitian Hamiltonians with a real
spectrum, and the conjecture on the connection between the reality of the
spectrum and the PT-invariance of these Hamiltonians due to Bender and
Boettcher$^{1}$, have given rise to a growing interest in the literature$%
^{2} $. Indeed, the above-mentioned Hamiltonians form a subclass of the
class of \textquotedblleft pseudo-Hermitian\textquotedblright\ operators, 
\textit{i.e.}, those operators which satisfy the equation 
\begin{equation}
A^{\dagger }=\eta A\eta ^{-1}
\end{equation}%
with 
\begin{equation}
\eta =\eta ^{\dagger }.
\end{equation}

Pseudo-Hermitian operators were introduced in the early 40's by Dirac$^{3}$
and Pauli$^{4}$ in order to overcome some divergence difficulties in physics
by using an indefinite metric associated with $\eta $ , and were later
resumed by Lee and Wick$^{5}$ (who firstly, at the best of our knowledge,
used the term \textquotedblleft pseudo-Hermiticity\textquotedblright ). More
recently, many interesting properties of such operators have been examined
and their spectrum has been suitably characterized.$^{6,7}$

We aim in the present paper to point out further properties of
pseudo-Hermitian operators that are relevant from a physical viewpoint. To
this end, we introduce in Sect. II the possibly broader class of \textit{%
weakly} pseudo-Hermitian operators, \textit{i.e.}, those operators which
satisfy Eq. (1) without any constraint on the (linear, invertible) operator $%
\eta $ , and show that, whenever one considers only diagonalizable
operators, this class actually coincides with the class of all the
pseudo-Hermitian operators. Hence the condition in Eq. (2) can be dropped
when defining (diagonalizable) pseudo-Hermitian operators, which is useful
from several viewpoints (in particular, it simplifies checking Eq. (1)).
Moreover, we show in Sect. III that a diagonalizable operator $H$ is
(weakly) pseudo-Hermitian if and only if an antilinear involutory operator
exists which commutes with it. This result has a number of relevant
consequences; in particular, in every theory which admits a time-reversal
invariance, or a CPT-invariance, the Hamiltonian must necessarily be a
(weakly) pseudo-Hermitian operator. Furthermore, the above result is
strictly intertwined with an old theorem$^{8}$ of group representation
theory, according to which a set of operators admits an involutory
antilinear mapping that commutes with it if and only if all the operators in
the set can assume conjointly a real form in a suitable basis. Indeed, by
using this theorem together with the above results, we conclude in Sect. IV
that for any diagonalizable (weakly) pseudo-Hermitian operator $H$ a basis
exists in which $H$ has a real form. If this basis coincides with the
eigenbasis of $H$, then $H$ also has a real spectrum. Finally, we illustrate
our results by means of an example, considering the special case of the
complex Morse potential$^{9}$ in Sect. V.

\section{The spectra of weakly pseudo-Hermitian operators}

As we wrote in the Introduction, we introduce here a new class of operators,
whose properties will be studied in the following.

\textbf{Definition 1.} \textit{A linear operator }$A$\textit{\ is said to be 
}weakly\textit{\ }pseudo-Hermitian\textit{\ if a linear, invertible operator 
}$\eta $\textit{\ exists such that}\ \ 
\begin{equation}
\eta A\eta ^{-1}=A^{\dagger }.  \tag{1}
\end{equation}

The above definition generalizes the definition of pseudo-Hermitian
operators since we do not assume $\eta =\eta ^{\dagger }$ as is required in
the standard definition of pseudo-Hermitian operators$^{5,6}$.

As in Refs. 6 and 7, we consider here only diagonalizable operators;
moreover, for the sake of simplicity, we consider only discrete spectra (see
however Sect. V, where a potential with a continuous spectrum is explicitly
studied). Whenever $H$ is a diagonalizable operator with a discrete
spectrum, a complete biorthonormal eigenbasis $\left\{ \left\vert \psi
_{n},a\rangle ,\right\vert \phi _{n},a\rangle \right\} $ exists$^{11}$, 
\textit{i.e.}, a basis such that 
\begin{equation}
H\left\vert \psi _{n},a\right\rangle =E_{n}\left\vert \psi
_{n},a\right\rangle ,\qquad H^{\dagger }\left\vert \phi _{n},a\right\rangle
=E_{n}^{\ast }\left\vert \phi _{n},a\right\rangle 
\end{equation}

\begin{equation}
\left\langle \phi _{m},b\right. \left\vert \psi _{n},a\right\rangle =\delta
_{mn}\delta _{ab}
\end{equation}%
\begin{equation}
\sum_{n}\sum_{a=1}^{d_{n}}\left\vert \phi _{n},a\right\rangle \left\langle
\psi _{n},a\right\vert =\sum_{n}\sum_{a=1}^{d_{n}}\left\vert \psi
_{n},a\right\rangle \left\langle \phi _{n},a\right\vert =\mathbf{1}
\end{equation}%
where $d_{n}$ denotes the degeneracy of $E_{n}$ , and $a$ and $b$ are
degeneracy labels.

The operator $H$ can then be written in the form%
\begin{equation*}
H=\sum_{n}\sum_{a=1}^{d_{n}}\left\vert \psi _{n},a\right\rangle
E_{n}\left\langle \phi _{n},a\right\vert .
\end{equation*}%
For the sake of brevity we also write the above basis $\left\{ \left\vert
\psi _{m}\rangle ,\right\vert \phi _{m}\rangle \right\} $ in the following,
with an obvious meaning of symbols. Then $H$ can also be written in the form%
\begin{equation*}
H=\sum_{m}\left\vert \psi _{m}\right\rangle E_{m}\left\langle \phi
_{m}\right\vert 
\end{equation*}%
(where it may occur that $E_{m}=E_{m^{\prime }}$ even if $m\neq m^{\prime }$%
). Furthermore, if $\left\{ u_{m}\right\} $ is any complete, orthonormal
basis in our space, we put in the following 
\begin{equation}
O=\sum_{m}\left\vert \psi _{m}\right\rangle \left\langle u_{m}\right\vert 
\text{ }
\end{equation}%
\ By using Eq. (6), we get%
\begin{equation*}
O^{-1}=\sum_{m}\left\vert u_{m}\right\rangle \left\langle \phi
_{m}\right\vert 
\end{equation*}%
and 
\begin{equation*}
O^{-1}HO=\sum_{m}\left\vert u_{m}\right\rangle \left\langle \phi
_{m}\right\vert \sum_{m^{\prime }}\left\vert \psi _{m^{\prime
}}\right\rangle E_{m^{\prime }}\left\langle \phi _{m^{\prime }}\right\vert
\sum_{m^{\prime \prime }}\left\vert \psi _{m^{\prime \prime }}\right\rangle
\left\langle u_{m^{\prime \prime }}\right\vert =\sum_{m}\left\vert
u_{m}\right\rangle E_{m}\left\langle u_{m}\right\vert .
\end{equation*}%
Moreover, 
\begin{equation}
(OO^{\dagger })H^{\dagger }(OO^{\dagger })^{-1}=\sum_{m}\left\vert \psi
_{m}\right\rangle E_{m}^{\ast }\left\langle \phi _{m}\right\vert .
\end{equation}

We can now state the following proposition, which embodies some results in
Ref. 6 on pseudo-Hermitian operators.

\textbf{Proposition \ 1.} \ \textit{Let }$\mathit{H}$\textit{\ be a
diagonalizable operator with a discrete spectrum. Then, the following
conditions are equivalent:}

\textit{i) $H$ is weakly pseudo-Hermitian;}

\textit{ii) the eigenvalues of $H$ occur in complex conjugate pairs, and for
each pair the multiplicities of both the eigenvalues are the same;}

\textit{iii) }$H$\textit{\ is pseudo-Hermitian.}

\textbf{Proof}. The implication $iii)\Rightarrow ii)$ is proven in Prop. 7
of Ref. 6. By observing that only the invertibility of $\eta $ is used in
this proof, in order to show that $\eta ^{-1}$ maps the eigensubspace of $%
H^{\dagger }$  associated with $E_{n}$ to that\ of $H$ associated with $%
E_{n}^{\ast }$ , and both the subspaces have the same dimension, one
immediately transforms this proof into a proof of the implication $%
i)\Rightarrow ii).$

The implication $ii)\Rightarrow iii)$ is also proven in Ref. 6. We provide
here, however, a somewhat different proof of it, which produces a useful
decomposition of $\eta $ (see Eq. (11)).

Let us therefore assume that condition $ii)$ holds, and use (whenever it is
necessary) the subscript `$_{0}$' to denote real eigenvalues and the
corresponding eigenvectors, and the subscript `$_{\pm }$' to denote the
complex eigenvalues with positive or negative imaginary part, respectively,
and the corresponding eigenvectors.

Then, let us consider the involutory operator $T$, defined as follows: 
\begin{equation}
T\left\vert \psi _{n_{\pm }},a\right\rangle =\left\vert \psi _{n_{\mp
}},a\right\rangle \text{ \quad (hence, }T\left\vert \psi
_{n_{0}},a\right\rangle =\left\vert \psi _{n_{0}},a\right\rangle \text{)}.
\end{equation}%
The explicit form of $T$ is 
\begin{eqnarray*}
T &=&T(\sum_{n_{0}}\sum_{a=1}^{d_{n_{0}}}\left\vert \psi
_{n_{0}},a\right\rangle \left\langle \phi _{n_{0}},a\right\vert
+\sum_{n_{+}}\sum_{a=1}^{d_{n_{+}}}\left\vert \psi _{n_{+}},a\right\rangle
\left\langle \phi _{n_{+}},a\right\vert
+\sum_{n_{-}}\sum_{a=1}^{d_{n_{-}}}\left\vert \psi _{n_{-}},a\right\rangle
\left\langle \phi _{n_{-}},a\right\vert )= \\
&&\sum_{n_{0},a}\left\vert \psi _{n_{0}},a\right\rangle \left\langle \phi
_{n_{0}},a\right\vert +\sum_{n_{+},n_{-},a}(\left\vert \psi
_{n_{-}},a\right\rangle \left\langle \phi _{n_{+}},a\right\vert +\left\vert
\psi _{n_{+}},a\right\rangle \left\langle \phi _{n_{-}},a\right\vert ).
\end{eqnarray*}%
The action of $T$ on the bras $\left\langle \phi _{n_{\pm }},a\right\vert $
easily follows from the expression above: 
\begin{equation}
\left\langle \phi _{n_{\pm }},a\right\vert T=\left\langle \phi _{n_{\mp
}},a\right\vert \qquad \text{(hence, }\left\langle \phi
_{n_{0}},a\right\vert T=\left\langle \phi _{n_{0}},a\right\vert \text{)}.
\end{equation}

Then, by simple calculations, one has 
\begin{equation}
THT=\sum_{m}\left\vert \psi _{m}\right\rangle E_{m}^{\ast }\left\langle \phi
_{m}\right\vert ,
\end{equation}%
and finally, comparing Eqs. (7) and (10), it follows 
\begin{equation*}
THT=(OO^{\dagger })H^{\dagger }(OO^{\dagger })^{-1},
\end{equation*}
hence condition $iii)$ follows at once, with 
\begin{equation}
\eta =(OO^{\dagger })^{-1}T=\sum_{n_{0},a}\left\vert \phi
_{n_{0}},a\right\rangle \left\langle \phi _{n_{0}},a\right\vert
+\sum_{n_{+},n_{-},a}(\left\vert \phi _{n_{+}},a\right\rangle \left\langle
\phi _{n_{-}},a\right\vert +\left\vert \phi _{n_{-}},a\right\rangle
\left\langle \phi _{n_{+}},a\right\vert )=\eta ^{\dagger }.
\end{equation}%
Finally, the proof of the Proposition can be completed by observing that the
implication $iii)\Rightarrow i)$ is obvious. $\blacksquare $

The introduction of the operator $T$ on the above proof and the
decomposition $\eta =(OO^{\dagger })^{-1}T$ in Eq. (11) allows one to obtain
immediately the characterization of the case of real spectrum. Indeed,
noting that $T=\mathbf{1}$ if and only if the spectrum of $H$ is real, the
following statement holds (see also the Theorem in Ref. 7).

\textbf{Proposition 2. }\textit{The spectrum of a diagonalizable weakly
pseudo-Hermitian (hence, of a diagonalizable pseudo-Hermitian) operator }$%
\mathit{H}$\textit{\ is real if and only if an operator }$\eta $\textit{\
exists such that }$\eta =(OO^{\dagger })^{-1}.$

Furthermore, the existence of an Hermitian operator $\eta $ whenever $H$ is
weakly pseudo-Hermitian implies that also in this case one can introduce an
Hermitian, indefinite inner product$^{4-6,10}$ which is invariant under the
time translation generated by $H$.

\section{Weakly pseudo-Hermitian operators and antilinear symmetries}

In order to discuss properly the next argument, we state the following
definition.

\textbf{Definition} \ \textbf{2}$^{5}$.\textit{\ Given the biorthonormal
basis }$\mathfrak{E}=\left\{ \left\vert \psi _{m}\rangle ,\right\vert \phi
_{m}\rangle \right\} $\textit{\ in a Hilbert space, we call }conjugation%
\textit{\ associated with it the involutory antilinear operator} \ \ 
\begin{equation}
\Theta _{\mathfrak{E}}=\sum_{m}\left\vert \psi _{m}\right\rangle
K\left\langle \phi _{m}\right\vert ,
\end{equation}%
\textit{where the operator K acts transforming each complex number on the
right into its complex conjugate.}

Let us discuss now the connection between the (weak) pseudo-Hermiticity
condition and the \textit{antilinear commutant}$^{8}$ of $H$ (\textit{i.e.},
the set of the antilinear, invertible operators which commute with it). This
connection was already acknowledged in Ref. 7, where the author shows that,
if $H$ commutes with an antilinear operator $A$, then condition $ii)$ of
Proposition 1 holds, and that a Hamiltonian\ with an antilinear symmetry $A$
has a real spectrum if and only if the symmetry is \textit{exact}$^{7}$(%
\textit{i.e.}, its eigenvectors are invariant under the action of $A$). The
latter statement can be rephrased, using Definition 2, by saying that 
\textit{the spectrum of $H$ is real if and only if }$\left[ H,\Theta _{%
\mathfrak{E}}\right] =0$ .

However, the above results enlighten only partially the key role of the
antilinear commutant of $H$ . Indeed, the following, more complete
proposition holds.

\textbf{Proposition 3}. \ \textit{Let $H$ be a diagonalizable operator with
a discrete spectrum. Then, the following conditions are equivalent:}

\textit{i)} \textit{an antilinear, invertible operator }$\Omega $\textit{\
exists such that }$\left[ H,\Omega \right] =0$ ;

\textit{ii) }$H$\textit{\ is (weakly) pseudo-Hermitian;}

\textit{iii)} \textit{an antilinear, involutory operator }$\hat{\Omega}$%
\textit{\ exists such that }$\left[ H,\hat{\Omega}\right] =0$ .

\textbf{Proof.} $i)\Rightarrow ii).$ Let $\Omega $ exist. Then, the linear
operator 
\begin{equation*}
\eta =(OO^{\dagger })^{-1}\Theta _{\mathfrak{E}}\Omega
\end{equation*}%
(where $\mathfrak{E}$ is the biorthonormal basis associated with $H$, and $O$
and $\Theta _{\mathfrak{E}}$ are defined as in Eqs. (6) and (12),
respectively) fulfils the condition stated by Eq. (1), hence $H$ is (weakly)
pseudo-Hermitian. Indeed, 
\begin{equation}
\Theta _{\mathfrak{E}}H\Theta _{\mathfrak{E}}^{-1}=\Theta _{\mathfrak{E}%
}H\Theta _{\mathfrak{E}}=\sum_{m}\left\vert \psi _{m}\right\rangle
E_{m}^{\ast }\left\langle \phi _{m}\right\vert ,
\end{equation}%
so that, recalling Eq. (7), 
\begin{equation*}
\eta H\eta ^{-1}=(OO^{\dagger })^{-1}\Theta _{\mathfrak{E}}\Omega H\Omega
^{-1}\Theta _{\mathfrak{E}}(OO^{\dagger })=(OO^{\dagger })^{-1}\Theta _{%
\mathfrak{E}}H\Theta _{\mathfrak{E}}(OO^{\dagger })=H^{\dagger }.
\end{equation*}
$ii)\Rightarrow iii).$ If $H$ is (weakly) pseudo-Hermitian, the eigenvalues
of $H$ occur in complex conjugate pairs, and for each pair the
multiplicities of both the eigenvalues are the same (Proposition 1). Then,
one easily sees, recalling the definition of the operator $T$ provided in
the proof of Proposition 1 and Eq. (10), that 
\begin{equation*}
\Theta _{\mathfrak{E}}H\Theta _{\mathfrak{E}}=THT.
\end{equation*}%
Hence the antilinear operator 
\begin{equation}
\hat{\Omega}=\Theta _{\mathfrak{E}}T=\sum_{n_{0},a}\left\vert \psi
_{n_{0},a}\right\rangle K\left\langle \phi _{n_{0},a}\right\vert
+\sum_{n_{+},n_{-},a}(\left\vert \psi _{n_{+},a}\right\rangle K\left\langle
\phi _{n_{-},a}\right\vert +\left\vert \psi _{n_{-},a}\right\rangle
K\left\langle \phi _{n_{+},a}\right\vert )
\end{equation}%
commutes with $H$. Finally, $\hat{\Omega}$ is involutory, (i. e., $\hat{%
\Omega}^{2}=\mathbf{1}$) as one can immediately verify by using the explicit
expression of $\hat{\Omega}$ in Eq. (14).

$iii)\Rightarrow i).$Obvious. $\blacksquare $

Proposition 3 has an interesting physical interpretation, as we have
emphasized in the Introduction. Indeed, whenever $H$ is the Hamiltonian of
some physical system, it establishes a link between the properties of $H$
(and of its spectrum) and the symmetries of the physical system described by
it. For, the time-reversal symmetry is associated, in complex quantum
mechanics, with an antilinear operator. Hence, whenever a physical system
admits such a symmetry (or else, more generally, it is invariant under the
combined action of the time-reversal operator times a linear one) the
antilinear commutant of its Hamiltonian must be non-void, hence $H$ is a
(weakly) pseudo-Hermitian operator. Vice versa any (weakly) pseudo-Hermitian
Hamiltonian admits an antilinear (involutory) symmetry.

Finally, since in the case of real spectrum the operator $T$ defined in the
proof of Proposition 1 is such that $T=\mathbf{1}$ , hence $\hat{\Omega}%
=\Theta _{\mathfrak{E}}$ , one obtains the following proposition.

\textbf{Proposition 4}.\textit{\ A diagonalizable, weakly pseudo-Hermitian
operator }$H$\textit{\ has a real spectrum if and only if it commutes with
the conjugation associated with its eigenbasis.}

\textbf{Remark} . While we were writing the final version of this paper,
some similar results have been obtained by other authors$^{12}$ (in
particular, having in mind the equivalence $i)\Leftrightarrow iii)$ in our
Proposition 1, Theorem 2 of Ref. 12 essentially states the equivalence $%
i)\Leftrightarrow ii)$ of our Proposition 3). Nevertheless, our presentation
is rather different and embodies the new condition $iii)$ in Proposition 3,
which has a number of interesting consequences, that we are going to explore
in the next section.

\section{Real form of the (weakly) pseudo-Hermitian operators}

According to Proposition 3, for any (weakly) pseudo-Hermitian operator $H$,
at least one involutory antilinear operator exists which commutes with it.
Then, it has been proven elsewhere$^{8}$ that any involutory antilinear
operator $\hat{\Omega}$ is a conjugation in some suitable basis; moreover,
in the basis associated with $\hat{\Omega}$, any operator commuting with $%
\hat{\Omega}$ has a real form.

The proof of this latter statement can be sketched as follows. If we denote
by $S$ the linear part of $\hat{\Omega}$ , \textit{i.e.}, $\hat{\Omega}=SK$
\ (where $K$ is the complex conjugation operator, see Sect. III), then $\hat{%
\Omega}^{2}=\mathbf{1}$ implies $SS^{\ast }=\mathbf{1}$ and this is possible
if and only if an $U$ exists such that $S=UU^{\ast -1}$. Then $\left[ H,\hat{%
\Omega}\right] =0$ implies $HUU^{\ast -1}=UU^{\ast -1}H^{\ast }$ , hence 
\begin{equation*}
U^{-1}HU=(U^{\ast -1}H^{\ast }U^{\ast })=(U^{-1}HU)^{\ast }.
\end{equation*}

Referring to the notation introduced in the present paper, let \ $\mathfrak{F%
}=\left\{ \left\vert v_{m}\right\rangle ,\left\vert w_{m}\right\rangle
\right\} $ be the biorthonormal basis associated, in the above sense, to the
conjugation $\hat{\Omega}$ which commutes with $H$ (of course, it may be an
orthonormal basis which occurs if and only if, for all $m$, $\left\vert
v_{m}\right\rangle =\left\vert w_{m}\right\rangle $ ), and let us consider
the matrix elements of $H$ in such basis. It is easy to verify that they are
real; indeed, on one hand, 
\begin{equation*}
\left\langle w_{i}\right\vert H\left\vert v_{k}\right\rangle =\left\langle
w_{i}\right\vert \sum_{n}\left\vert \psi _{n}\right\rangle E_{n}\left\langle
\phi _{n}\right\vert \left. v_{k}\right\rangle 
\end{equation*}%
and, on the other hand 
\begin{equation*}
\left\langle w_{i}\right\vert \hat{\Omega}H\hat{\Omega}\left\vert
v_{k}\right\rangle =\left\langle w_{i}\right\vert \sum_{m}\left\vert
v_{m}\right\rangle K\left\langle w_{m}\right\vert \sum_{n}\left\vert \psi
_{n}\right\rangle E_{n}\left\langle \phi _{n}\right\vert \sum_{m^{\prime
}}\left\vert v_{m^{\prime }}\right\rangle K\left\langle w_{m^{\prime
}}\right. \left\vert v_{k}\right\rangle =
\end{equation*}%
\begin{equation*}
\sum_{m,m^{\prime },n}\delta _{im}K\left\langle w_{m}\right. \left\vert \psi
_{n}\right\rangle E_{n}\left\langle \phi _{n}\right. \left\vert v_{m^{\prime
}}\right\rangle K\delta _{m^{\prime },k}=
\end{equation*}%
\begin{equation*}
\left\langle v_{k}\right\vert \sum_{n}\left\vert \phi _{n}\right\rangle
E_{n}^{\ast }\left\langle \psi _{n}\right. \left\vert w_{i}\right\rangle
=\left( \left\langle w_{i}\right\vert \sum_{n}\left\vert \psi
_{n}\right\rangle E_{n}\left\langle \phi _{n}\right. \left\vert
v_{k}\right\rangle \right) ^{\ast }.
\end{equation*}

Since, trivially, every operator which assumes a real form in some basis $%
\mathfrak{B}$ commutes with the conjugation associated with $\mathfrak{B}$ ,
we have thus proven the following proposition.

\textbf{Proposition 5}. \textit{An operator }$H$\textit{\ is (weakly)
pseudo-Hermitian if and only if a basis exists in which it assumes a real
form.}

The results in Propositions 1, 3 and 5 can be collected together, obtaining
a set of six equivalent conditions which can be useful in order to
characterize the Hamiltonians that we are considering. In particular, the
statement in Proposition 5 can be used to write a (weakly) pseudo-Hermitian
operator in a more manageable form (an example of basis transformation which
puts a particular Hamiltonian in real form is in the next Section).

\section{An example: the complex Morse potential}

Let us verify the results obtained in the previous sections in the special
case of the complex Morse potential$^{9}$. This was extensively studied, for
instance, in Ref. 9 and its spectrum was predicted to be real by means of
group theoretic techniques$^{13}$.

The Morse potential is given by 
\begin{equation}
V(x)=(A+iB)^{2}e^{-2x}-(A+iB)(2C+1)e^{-x}\qquad (A,B,C\in \mathbf{R}).
\end{equation}
Putting $\rho =\sqrt{A^{2}+B^{2}},\theta =\arctan \frac{2B}{A},k=2C+1,$ we
get%
\begin{equation}
V(x)=\rho ^{2}e^{-2x+i\theta }-k\rho e^{-x+i\theta /2}.
\end{equation}

Following Ref. 9, let us introduce the (Hermitian) operator $e^{-\theta
p}\quad (\theta \in \mathbf{R},\quad p=-i\frac{d}{dx})$ . Hence, we obtain
the following Equation$^{9}$ 
\begin{equation}
e^{-\theta p}V(x)e^{\theta p}=V(x+i\theta )=V^{\ast }(x).
\end{equation}%
This equation shows that $V$ is a pseudo-Hermitian (but non PT-symmetric)
operator. Let us put now $\hat{\Omega}=SK=e^{\theta p}K$ . By using Eq. (17)
one gets 
\begin{equation*}
\hat{\Omega}V=V\hat{\Omega},
\end{equation*}%
which agrees with Proposition 3. Then, a straightforward calculation shows
that $S^{\ast }=e^{-\theta p}=S^{-1}$, hence $\hat{\Omega}$ is involutory,
which also agrees with Proposition 3. Moreover, one gets by inspection that 
\begin{equation*}
S=e^{\theta p/2}(e^{\theta p/2})^{\ast -1}=UU^{\ast -1}.
\end{equation*}%
Thus, finally, 
\begin{equation*}
U^{-1}VU=e^{-\theta p/2}Ve^{\theta p/2}=V(x+i\theta /2)=\rho
^{2}e^{-2x}-k\rho e^{-x}=(U^{-1}VU)^{\ast }
\end{equation*}%
which agrees with Proposition 5.

\bigskip

\bigskip

\bigskip

{\large References}

\bigskip

$^{1}$ C. M. Bender and S. Boettcher, \textit{Phys. Rev. Lett}. \textbf{80}
(1998) 5243.

$^{2}$ See, for instance, M. Znojil, ''Should PT symmetric quantum mechanics
be interpreted as nonlinear?'', arXiv: quant-ph/0103054 (2001), and
references therein.

$^{3}$ P. A. M. Dirac,\textit{\ Proc. Roy. Soc}. \textbf{A180} (1942) 1.

$^{4}$ W. Pauli, \textit{Rev. Mod.. Phys}. \textbf{15} (1943) 175.

$^{5}$ T. D. Lee and G. C. Wick, \textit{Nucl. Phys.} \textbf{B9} (1969) 209.

$^{6}$ A. Mostafazadeh, \textit{Journ. Math. Phys}. \textbf{43} (2002) 205.

$^{7}$ A. Mostafazadeh, ''Pseudo-Hermiticity versus PT-Symmetry II'', arXiv:
math-ph/0110016 (2001).

$^{8}$ R. Ascoli, C. Garola, L. Solombrino and G. C. Teppati, ''Real versus
complex representations and linear-antilinear commutant'', in \textit{%
Enz/Mehra (eds.), Physical Reality and Mathematical Description }(D. Reidel
Publishing Company, Dordrecht-Holland, 1974), 239.

$^{9}$ Z. Ahmed,''Pseudo-Hermiticity of Hamiltonians under imaginary shift
of the coordinate: real spectrum of complex potentials'', arXiv:
quant-ph/0108016 (2001).

$^{10}$ G. S.Japaridze, ''Space of state vectors in PT symmetrized quantum
mechanics'', arXiv: quant-ph/0104077 (2001) .

$^{11}$ F. H. M. Faisal and J. V. Moloney, \textit{J. Phys. B: At. Mol. Phys}%
. \textbf{14} (1981) 3603.

$^{12}$A. Mostafazadeh, \textquotedblright Pseudo-Hermiticity versus
PT-Symmetry III\textquotedblright , arXiv: math-ph/0203005 (2002).

$^{13}$B. Bagchi and C. Quesne, \textit{Phys. Lett}. \textbf{A 273} (2000)
285.

\end{document}